\documentclass[journal]{IEEEtran}
\usepackage{amsmath}
\usepackage{cite}
\usepackage{graphicx}
\usepackage{epstopdf}
\usepackage{amsfonts,amsmath,amssymb}
\usepackage{graphicx}
\usepackage{url}
\usepackage{bm}
\usepackage{bbm}
\usepackage{subfigure}
\usepackage{stfloats}
\usepackage{color}

\usepackage{algorithmic}
\usepackage{algorithm}

\DeclareMathOperator*{\argmax}{argmax}
\DeclareMathOperator*{\argmin}{argmin}

\newtheorem{corollary}{\textbf{Corollary}}
\newtheorem{theorem}{\textbf{Theorem}}
\newtheorem{proposition}{\textbf{Proposition}}

\newcommand{\Hnull}{\mathcal{H}_0}
\newcommand{\Halt}{\mathcal{H}_1}

\newcommand{\Honull}{\mathcal{{D}}_0}
\newcommand{\Hoalt}{\mathcal{{D}}_1}

\begin{document}

\title{Gaussian Signalling for Covert Communications}

\author{

Shihao~Yan, \IEEEmembership{Member, IEEE,} Yirui Cong, \IEEEmembership{Member, IEEE,} Stephen V. Hanly, \IEEEmembership{Fellow, IEEE,}\\ and Xiangyun~Zhou, \IEEEmembership{Senior Member, IEEE}

\thanks{S. Yan and S. V. Hanly are with the School of Engineering, Macquarie University, Sydney, NSW 2109, Australia
(e-mails: \{shihao.yan, stephen.hanly\}@mq.edu.au).}
\thanks{Y. Cong is with the College of Intelligence Science and Technology, National University of Defense Technology, Changsha, Hunan 410073, China (e-mail: congyirui11@nudt.edu.cn).}
\thanks{X. Zhou is with Research School of Electrical, Energy and Materials Engineering, Australian National University, Canberra, ACT 2601, Australia (e-mail: xiangyun.zhou@anu.edu.au).}
\thanks{This research was supported by Macquarie University under the MQRF Fellowship and by
the CSIRO Macquarie University Chair in Wireless Communications. This Chair has been established with funding provided
by the Science and Industry Endowment Fund. This work was also partially supported by the National Natural Science Foundation of China under Grant 61801494.}
}

\markboth{IEEE Transactions on Wireless Communications}{Yan \MakeLowercase{\textit{et al.}}: Gaussian Signalling for Covert Communications}

\maketitle

\vspace{-2cm}

\begin{abstract}
In this work, we examine the optimality of Gaussian signalling for covert communications with an upper bound on $\mathcal{D}(p_{_1}||p_{_0})$ or $\mathcal{D}(p_{_0}||p_{_1})$ as the covertness constraint, where $\mathcal{D}(p_{_1}||p_{_0})$ and $\mathcal{D}(p_{_0}||p_{_1})$ are different due to the asymmetry of Kullback-Leibler divergence, $p_{_0}(y)$ and $p_{_1}(y)$ are the likelihood functions of the observation ${y}$ at the warden under the null hypothesis (no covert transmission) and alternative hypothesis (a covert transmission occurs), respectively. Considering additive white Gaussian noise at both the receiver and the warden, we prove that Gaussian signalling is optimal in terms of maximizing the mutual information of transmitted and received signals for covert communications with an upper bound on $\mathcal{D}(p_{_1}||p_{_0})$ as the constraint. More interestingly, we also prove that Gaussian signalling is not optimal for covert communications with an upper bound on $\mathcal{D}(p_{_0}||p_{_1})$ as the constraint, for which as we explicitly show skew-normal signalling can outperform Gaussian signalling in terms of achieving higher mutual information. {Finally, we prove that, for Gaussian signalling, an upper bound on $\mathcal{D}(p_{_1}||p_{_0})$ is a tighter covertness constraint in terms of leading to lower mutual information than the same upper bound on $\mathcal{D}(p_{_0}||p_{_1})$, by proving $\mathcal{D}(p_{_0}||p_{_1}) \leq \mathcal{D}(p_{_1}||p_{_0})$.}
\end{abstract}



\section{Introduction}

With Internet of Things (IoT) coming to reality, people and organizations become more dependent on wireless devices to share private information (e.g., location information, physiological information for e-health).
As a consequence, there are increasing concerns on security and privacy in such applications. Against this background, physical layer security has been widely used to address and enhance wireless communication security, which is compatible and complementary to traditional cryptographic techniques \cite{bloch2011physical,duong2016trusted}. However, although physical layer security can protect the content of wireless communications \cite{bloch2011physical,duong2016trusted}, it cannot fully address privacy concerns. For example, the exposure of a wireless transmission may disclose a user's location information, which may violate the privacy of the user and this cannot be resolved by physical layer security or cryptographic techniques. Against this background, covert communication is emerging as a new technique to achieve a strong security and privacy in wireless communications (i.e., hiding wireless transmissions) \cite{bash2013limits,che2014reliable,lee2015achieving,bash2015hiding}.

Hiding wireless transmissions was only partially addressed by spread spectrum, which focuses on hiding military wireless transmissions by spreading transmit power to make it appear noise like~\cite{simon1994spread}. However, the achieved covertness by spread spectrum has never been proven theoretically, because there is no fundamental understanding on when or how often spread spectrum fails to hide wireless transmissions. As such, recent cutting-edge research on wireless communication security has focused on the fundamental limits of covert communications (e.g., \cite{bash2013limits,bloch2016covert,wang2016fundamental,wu2017limits}). In covert communications, a transmitter (Alice) desires to transmit information to a legitimate receiver (Bob) without being detected by a warden (Willie), who is collecting observations to detect this transmission. Considering additive white Gaussian noise (AWGN) channels, a square root law was established in\cite{bash2013limits}, which states that Alice can transmit no more than $\mathcal{O}(\sqrt{n})$ bits in $n$ channel uses covertly and reliably to Bob. Besides, some works in the literature focused on the design and performance analysis of covert communications in practical application scenarios, for example, by considering unknown background noise power \cite{goeckel2016covert}, ignorance of transmission time \cite{bash2016covert}, noise uncertainty \cite{he2017on}, delay constraints \cite{yan2017covert,yan2018delay}, channel uncertainty \cite{khurram2017covert}, practical modulation \cite{bloch2017optimal}, uninformed jamming \cite{sobers2017covert}, relay networks\cite{hu2017covertGC,hu2018covertTWC}, broadcast channels \cite{arumugam2017covert}, key generation \cite{tah2017covert}, and artificial noise \cite{soltani2014covert,soltani2014covert1}.

In covert communications, for an optimal detector at Willie, we have $\xi^{\ast} = 1 - \mathcal{V}_T(p_{_0}, p_{_1})$, where $\xi^{\ast}$ is the minimum detection error probability and $\mathcal{V}_T(p_{_0}, p_{_1})$ is the total variation between the likelihood function $p_{_0}(y)$ of the observation $y$ under the null hypothesis (when Alice does transmit to Bob) and the likelihood function $p_{_0}(y)$ under the alternative hypothesis (when Alice transmits to Bob). Due to the mathematically intractable expressions for $\mathcal{V}_T(p_{_0}, p_{_1})$, Kullback-Leibler (KL) divergence (i.e., relative entropy) has been widely adopted to limit the detection performance at Willie in the literature of covert communications. Specifically, as per the Pinsker's inequality we have $\mathcal{V}_T(p_{_0}, p_{_1}) \leq \sqrt{\mathcal{D}(p_{_1}||p_{_0})/2}$ or $\mathcal{V}_T(p_{_0}, p_{_1}) \leq \sqrt{\mathcal{D}(p_{_0}||p_{_1})/2}$, where $\mathcal{D}(p_{_1}||p_{_0})$ is the KL divergence from $p_{_1}({y})$ to $p_{_0}({y})$ and $\mathcal{D}(p_{_0}||p_{_1})$ is the KL divergence from $p_{_0}({y})$ to $p_{_1}({y})$. Then, the covertness constraint $\xi^{\ast} \geq 1 - \epsilon$ can be guaranteed by two constraints on these KL divergences, i.e., $\mathcal{D}(p_{_1}||p_{_0}) \leq 2\epsilon^2$ and $\mathcal{D}(p_{_0}||p_{_1}) \leq 2\epsilon^2$, where $\epsilon$ is a small value determining the required covertness.
Based on the Pinsker's inequality as detailed above, the two constraints determined by the KL divergences are stricter than the covertness constraint $\xi^{\ast} \geq 1 - \epsilon$. This means that the covertness achieved under the former constraints (i.e., $\mathcal{D}(p_{_1}||p_{_0}) \leq 2\epsilon^2$ and $\mathcal{D}(p_{_0}||p_{_1}) \leq 2\epsilon^2$) can be achieved in practice under the later covertness constraint.  As such, the developed covert communication systems under the constraint $\mathcal{D}(p_{_1}||p_{_0}) \leq 2\epsilon^2$ or $\mathcal{D}(p_{_0}||p_{_1}) \leq 2\epsilon^2$ are fully operational in practice.

{We do not at present have any bound on the difference between optimality under actual covertness constraint and optimality under either of the KL constraints.
%
%
One reason for not obtaining such a bound is the seeming intractability of characterizing performance under the exact constraint.
Further work is required in this direction and we hope our work will provide motivation for obtaining better bounds on the actual detection error probability in future works.
The KL constraints that we use have been widely adopted in the literature on covert communications  (e.g.,[3,8,9,24]), and enable us to obtain analytical results of a conservative nature, which can be applied to solve network optimization problems in the context of covert communications.
}

A closely related research topic to covert communications is the stealth communication problem \cite{hou2014effective,song2018stealth}. The major difference between covert communications and stealth communications is that Alice does not transmit to Bob (i.e., ``zero symbols'' input) in the null hypothesis for covert communications, while Alice transmits non-zero symbols, following a non-zero innocent distribution,  to Bob in the null hypothesis for stealth communications \cite{hou2014effective,song2018stealth}.

In the literature of covert communications, these two specific constraints have been widely used in different works in the context of covert communications. {For example, with the aid of $\mathcal{D}(p_{_0}||p_{_1})$ to bound the detection error probability in part of the considered covertness constraint, the authors of \cite{bash2013limits} established the square root limit on the amount of information that can be transmitted from Alice to Willie reliably for any $\epsilon > 0$.} With the same constraint, the work \cite{wu2017limits} extended this square root law into a two-hop wireless system, where the source intends to communicate with
the destination covertly via a Decode-and-Forward relay node and the conducted analysis shows that this square root law can be extended into a multi-hop system.
Meanwhile, using $\mathcal{D}(p_{_1}||p_{_0}) \leq 2\epsilon^2$ as the covertness constraint, the authors of \cite{wang2016fundamental} proved that the square-root law holds for a broad class of discrete memoryless channels (DMCs), in which the scaling constant of the amount of information with respect to the square root of the blocklength has been determined for DMCs and AWGN channels. With the same covertness constraint, the shared key bits to guarantee the square-root law was quantified and the condition for which a secret key is not required was determined in \cite{bloch2016covert}. In addition, with the same constraint the authors of \cite{suria2016keyless,suria2018covert} extended the main results of \cite{bloch2016covert} into a discrete memoryless multiple-access channel, in which the pre-constant of the scaling is identified. Furthermore, with $\mathcal{D}(p_{_1}||p_{_0}) \leq 2\epsilon^2$ as the constraint, \cite{tahmasbi2016second,tahmasbi2017first} characterized the second order asymptotics of the number of bits that can be reliably and covertly transmitted and \cite{tahmasbi2017error} examined the error exponent of covert communications over binary-input discrete memoryless channels.

{We note that the aforementioned two KL divergences (i.e., $\mathcal{D}(p_{_1}||p_{_0})$ and $\mathcal{D}(p_{_0}||p_{_1})$) are different due to the asymmetric property of the KL divergence \cite{cover2002elements}. However, the resultant differences of using the two constraints, i.e., $\mathcal{D}(p_{_1}||p_{_0}) \leq 2\epsilon^2$ and $\mathcal{D}(p_{_0}||p_{_1}) \leq 2\epsilon^2$, in the context of covert communications have never been examined. This mainly motivates this work. We would like to clarify that the square root law was established with $\mathcal{D}(p_{_0}||p_{_1}) \leq 2\epsilon^2$ as the covertness constraint, while the result under the constraint $\mathcal{D}(p_{_1}||p_{_0}) \leq 2\epsilon^2$ has not been clarified.
As we will show in this work, using these two different constraints does affect the exact amount of covert information that can be reliably transmitted from Alice to Bob for a given value of $\epsilon$, although the difference becomes negligible as $\epsilon$ decreases to zero.}
{We note that in the literature Gaussian signalling was adopted with both $\mathcal{D}(p_{_1}||p_{_0}) \leq 2\epsilon^2$ and $\mathcal{D}(p_{_0}||p_{_1}) \leq 2\epsilon^2$ as constraints in covert communications, since Gaussian signalling at least can maximize the communication performance from Alice to Bob. However, the optimality of Gaussian signalling was not discussed under either of these two constraints.}
As we will show in this work, we have different signalling strategies to achieve the maximum amount of covert information subject to $\mathcal{D}(p_{_1}||p_{_0}) \leq 2\epsilon^2$ or to $\mathcal{D}(p_{_0}||p_{_1}) \leq 2\epsilon^2$. Considering AWGN at both Bob and Willie, the main contributions together with the key results of this work are summarized as below.
\begin{itemize}
\item We prove that Gaussian signalling is optimal in terms of maximizing the mutual information between the transmitted signal $\mathbf{x}$ sent by Alice and the signal $\mathbf{z}$ received by Bob subject to $\mathcal{D}(p_{_1}||p_{_0}) \leq 2\epsilon^2$.

\item We prove that Gaussian signalling is \textbf{not} optimal in terms of maximizing $I(\mathbf{x}; \mathbf{z})$ subject to $\mathcal{D}(p_{_0}||p_{_1}) \leq 2\epsilon^2$ for covert communications. We explicitly show that skew-normal signalling strategy can achieve a higher $I(\mathbf{x}; \mathbf{z})$ subject to $\mathcal{D}(p_{_0}||p_{_1}) \leq 2\epsilon^2$  than Gaussian signalling.


\item {We prove that Gaussian signalling minimizes the KL divergence $\mathcal{D}(p_{_1}||p_{_0})$ for any given average transmit power constraint on $\mathbf{x}$}, which explains why Gaussian signalling is optimal for covert communications with $\mathcal{D}(p_{_1}||p_{_0}) \leq 2\epsilon^2$ as the constraint, while Gaussian signalling cannot minimize the KL divergence $\mathcal{D}(p_{_0}||p_{_1})$.

\item We prove $\mathcal{D}(p_{_0}||p_{_1}) \leq \mathcal{D}(p_{_1}||p_{_0})$ for Gaussian signalling. This leads to the fact that $\mathcal{D}(p_{_0}||p_{_1})$ determines a tighter lower bound on Willie's actual minimum detection error probability $\xi^{\ast}$ than $\mathcal{D}(p_{_1}||p_{_0})$. An important implication is that the use of $\mathcal{D}(p_{_0}||p_{_1}) \leq 2\epsilon^2$ as the covert constraint gives a higher value of $I(\mathbf{x}; \mathbf{z})$.
\end{itemize}

The rest of this paper is organized as follows. Section \ref{system_model} details the system model and the focused problem of this work. Section~\ref{section_10} proves that Gaussian signalling is optimal for covert communications with $\mathcal{D}(p_{_1}||p_{_0}) \leq 2\epsilon^2$ as the constraint. Section~\ref{section_01} proves that Gaussian signalling is not optimal for covert communications with $\mathcal{D}(p_{_0}||p_{_1}) \leq 2\epsilon^2$ as the constraint. In Section~\ref{section_gaussian}, we examine the performance of covert communications with Gaussian signalling, where $\mathcal{D}(p_{_0}||p_{_1}) \leq \mathcal{D}(p_{_1}||p_{_0})$ is proved. Section~\ref{numerical_re} explicitly shows that skew-normal signalling is better than Gaussian signalling in terms of achieving a higher $I(\mathbf{x}; \mathbf{z})$ subject to $\mathcal{D}(p_{_0}||p_{_1}) \leq 2\epsilon^2$. Finally, Section \ref{conclusion} makes some concluding remarks.

\emph{Notation:} Given a random vector $\mathbf{x}$ and its realization $x$, $\mathbf{x}[i]$ and $x[i]$ denote the $i$-th element of $\mathbf{x}$ and $x$, respectively. The expectation operator is denoted by $\mathbb{E}[\cdot]$ and $\mathcal{N}(0,\sigma^2)$ denotes the normal distribution with zero mean and variance $\sigma^2$.


\section{System Model}\label{system_model}

\subsection{Channel Model}

\begin{figure}[!t]
    \begin{center}
        \includegraphics[width=0.7\columnwidth]{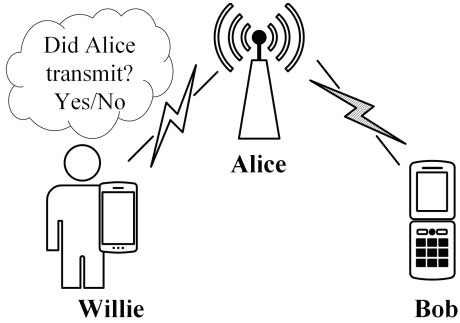}
        \caption{Illustration of the system model for covert communications.}
        \label{fig:system}
    \end{center}
\end{figure}

The system model for covert communications is illustrated in Fig.~\ref{fig:system}, where each of Alice, Bob, and Willie is equipped with a single antenna. We assume the channel from Alice to Bob and the channel from Alice to Willie are only subject to AWGN. In this work, we assume that Alice transmits one real-valued symbol $\mathbf{x}[i]$ to Bob in the $i$-th channel use, while Willie is passively collecting one observation on Alice's transmission to detect whether or not Alice has transmitted the signal to Bob.
We denote the AWGN at Bob and Willie in the $i$-th channel use as $\mathbf{n}_b[i]$ and $\mathbf{n}_w[i]$, respectively, where the elements of $\mathbf{n}_b$ or $\mathbf{n}_w$ are identically independently distributed (i.i.d.) and thus we have $\mathbf{n}_b[i] \sim \mathcal{N}(0,\sigma_b^2)$, $\mathbf{n}_w[i] \sim \mathcal{N}(0,\sigma_w^2)$, while $\sigma_b^2$ and $\sigma_w^2$ are the noise variances at Bob and Willie, respectively. In addition, we assume that $\mathbf{x}$, $\mathbf{n}_b$, and $\mathbf{n}_w$ are mutually independent and we the number of channel uses (denoted by $N$) is sufficient large such that the elements of $\mathbf{x}$ are i.i.d.. We further assume that Alice's transmit power of $\mathbf{x}[i]$ is fixed and denoted as $P_x$, i.e., we have $\mathbb{E}[|\mathbf{x}[i]|^2] = P_x$.

\subsection{Binary Hypothesis Testing at Willie}

In order to detect the presence of covert communications, Willie must distinguish between the following two hypotheses:
\begin{eqnarray}\label{hypothses}
 \left\{ \begin{aligned}\label{ncon}
        \ & \Hnull:~\mathbf{y}[i] = \mathbf{n}_w[i], \;\;\\
        \ & \Halt:~\mathbf{y}[i] = \mathbf{x}[i] + \mathbf{n}_w[i],
         \end{aligned} \right.
\end{eqnarray}
where $\Hnull$ denotes the null hypothesis where Alice has not transmitted signals, $\Halt$ denotes the alternative hypothesis where Alice has transmitted, and $\mathbf{y}[i]$ is the received signal at Willie in the $i$-th channel use.

In general, the detection error probability is adopted to measure Willie's detection performance, which is defined as
\begin{align}
\xi = \alpha + \beta,
\end{align}
where $\alpha \triangleq \Pr(\Hoalt|\Hnull)$ is the false positive rate, $\beta \triangleq \Pr(\Honull|\Halt)$ is the miss detection rate, and $\Hoalt$ and $\Honull$ are the binary decisions that infer whether Alice's transmission is present or not, respectively.
In covert communications, Willie's ultimate goal is to detect the presence of Alice's transmission with the minimum detection error probability $\xi^{\ast}$, which is achieved by using an optimal detector. Then, the covertness constraint can be written as $\xi^{\ast} \geq 1 - \epsilon$ for a given $\epsilon$, where the value of $\epsilon$ is predetermined and is normally small in order to guarantee sufficient covertness.

For an optimal detector at Willie, we have \cite{bash2013limits,lehmann2005testing,cover2002elements}
\begin{align}\label{distance}
\xi^{\ast} = 1 - \mathcal{V}_T(p_{_0}, p_{_1}) = 1 - \frac{1}{2}\|p_{_0}(y) - p_{_1}(y)\|_1,
\end{align}
where $\mathcal{V}_T(p_{_0}, p_{_1})$ is the total variation between $p_{_0}(y)$ and $p_{_0}(y)$, $\|a-b\|_1$ is the $\mathcal{L}_1$ norm, and $p_{_0}(y) = f(y|\Hnull)$ and $p_{_1}(y) = f(y|\Halt)$ are the likelihood functions of $\mathbf{y}$ under $\Hnull$ and $\Halt$, respectively. In general, computing $\mathcal{V}_T(p_{_0}, p_{_1})$ analytically is intractable and thus Pinsker's inequality is normally adopted to upper bound it. Based on Pinsker's inequality, we have
\begin{align}\label{pinsker1}
\mathcal{V}_T(p_{_0}, p_{_1}) \leq \sqrt{\frac{1}{2}\mathcal{D}(p_{_1}||p_{_0})},
\end{align}
or
\begin{align}\label{pinsker2}
\mathcal{V}_T(p_{_0}, p_{_1}) \leq \sqrt{\frac{1}{2}\mathcal{D}(p_{_0}||p_{_1})},
\end{align}
where $\mathcal{D}(p_{_1}||p_{_0})$ is the Kullback-Leibler (KL) divergence from $p_{_1}(y)$ to $p_{_0}(y)$, which is given by
\begin{align}\label{KL_10}
\mathcal{D}(p_{_1}||p_{_0}) = \int_{\mathcal{Y}} p_{_1}(y) \log\frac{p_{_1}(y)}{p_{_0}(y)} d {y},
\end{align}
and $\mathcal{D}(p_{_0}||p_{_1})$ is the KL divergence from $p_{_0}(y)$ to $p_{_1}(y)$, which is given by
\begin{align}\label{KL_qp}
\mathcal{D}(p_{_0}||p_{_1}) = \int_{\mathcal{Y}} p_{_0}(y) \log\frac{p_{_0}(y)}{p_{_1}(y)} d {y}.
\end{align}
We note that both \eqref{pinsker1} and \eqref{pinsker2} are valid, although they are different due to the asymmetry of the KL divergence, which can be seen from \eqref{KL_10} and \eqref{KL_qp}.

Following \eqref{distance} and \eqref{pinsker1}, it is sufficient to guarantee
\begin{align}\label{CR_pq}
\mathcal{D}(p_{_1}||p_{_0}) \leq 2\epsilon^2,
\end{align}
in order to guarantee $\xi^{\ast} \geq 1 - \epsilon$.
Alternatively, following \eqref{distance} and \eqref{pinsker2}, it is also sufficient to guarantee
 \begin{align}\label{CR_qp}
\mathcal{D}(p_{_0}||p_{_1}) \leq 2\epsilon^2,
\end{align}
in order to guarantee $\xi^{\ast} \geq 1 - \epsilon$.
{This is the main reason why Gaussian signaling is widely adopted in the literature of covert communications (e.g., \cite{bash2013limits,wu2017limits}), where we note that in \cite{bash2013limits} Gaussian signalling was used in the construction for the achievability result,} while \eqref{CR_qp} is also adopted (e.g., \cite{bloch2016covert,wang2016fundamental}).
We also note that these two constraints are both sufficient as per Pinsker's inequality. However, the difference between these two constraints in the context of covert communications has never been clarified. Noting that the elements of $\mathbf{y}$ are i.i.d., we have
\begin{align}
\mathcal{D}(p_{_1}||p_{_0}) &= N \times \mathcal{D}(p_{_1}(y[i])||p_{_0}(y[i])),\label{CR_pq_i}\\
\mathcal{D}(p_{_0}||p_{_1}) &= N \times \mathcal{D}(p_{_0}(y[i])||p_{_1}(y[i])),\label{CR_qp_i}
\end{align}
where we recall that $N$ is the total number of channel uses, {which is assumed to be sufficiently large in this work.}

\subsection{Mutual Information}

When Alice transmits $\mathbf{x}[i]$, the received signal at Bob in the $i$-th channel use is given by
\begin{align}\label{xn}
\mathbf{z}[i] = \mathbf{x}[i] + \mathbf{n}_b[i].
\end{align}
Then, the mutual information of $\mathbf{x}$ and $\mathbf{z}$ is given by
\begin{align}\label{mutual_N}
I(\mathbf{x};\mathbf{z}) = \int_{\mathcal{Z}}\int_{\mathcal{X}}p(x,z) \log \frac{p(x,z)}{p(x) p(z)} dx dz,
\end{align}
where $p(x,z)$ is the joint probability function of $\mathbf{x}$ and $\mathbf{z}$, $p(z)$ is the marginal probability distribution of $\mathbf{z}$, $\mathcal{Z}$ is the set of $\mathbf{z}$, and $\mathcal{X}$ is the set of $\mathbf{x}$.
For $\mathbf{n}_b[i] \sim \mathcal{N}(0,\sigma_b^2)$, $p(x[i]) = \mathcal{N}(0,P)$ can maximize $I(\mathbf{x};\mathbf{z})$ subject to $\mathbb{E}[|\mathbf{x}[i]|^2] = P$ as per \cite[Theorem 8.6.5]{cover2002elements}. {This is the main reason why Gaussian signaling is widely adopted in the literature of covert communications (e.g., \cite{bash2013limits,wu2017limits}).} Noting that the elements of $\mathbf{x}$ are i.i.d. and the elements of $\mathbf{z}$ are i.i.d., we have
\begin{align}\label{mutual_i}
I(\mathbf{x};\mathbf{z}) = N \times I(\mathbf{x}[i];\mathbf{z}[i]).
\end{align}

Considering \eqref{CR_pq_i}, \eqref{CR_qp_i}, and \eqref{mutual_i}, without loss of generality in this work we focus on one particular channel use, i.e., $\mathbf{x}$, $\mathbf{y}$, $\mathbf{z}$, $\mathbf{n}_b$, $\mathbf{n}_w$ and their realizations are $1$-dimensional in the rest of the paper.
%
%
As such, in the reminder of this work, we tackle whether Gaussian signalling is optimal in terms of maximizing $I(\mathbf{x},\mathbf{z})$ subject to different covertness constraints, i.e., $\xi^{\ast} \geq 1 - \epsilon$, $\mathcal{D}(p_{_1}||p_{_0}) \leq 2\epsilon^2$, and $\mathcal{D}(p_{_0}||p_{_1}) \leq 2\epsilon^2$.
%
%
Note that we do not use the rate defined in the limit of $N \rightarrow \infty$ as a performance metric in this work. This is due to the fact that this rate, as per the square root law, is zero regardless of the signalling strategy in covert communications, since the converse proof of the square root law is valid for an arbitrary signalling strategy [4]. Therefore, we cannot use this rate as an objective function to tackle the optimality of Gaussian signalling for covert communications.

\section{With $\mathcal{D}(p_{_1}||p_{_0}) \leq 2\epsilon^2$ as the Covertness Constraint}\label{section_10}

{In this section, we analytically prove that Gaussian signaling is optimal for covert communications in terms of maximizing $I(\mathbf{x},\mathbf{z})$ subject to $\mathcal{D}(p_{_1}||p_{_0}) \leq 2\epsilon^2$ and other related constraints. Mathematically, we prove the following theorem.}
\begin{theorem}\label{theorem1}
The zero-mean Gaussian signaling with variance $P_x^{\ast}$, i.e., $p(x) = \mathcal{N}(0,P_x^{\ast})$, is the solution to the following optimization problem
\begin{subequations}\label{opt10}
\begin{align}
 \argmax_{p(x),~P_x} ~~&I(\mathbf{x},\mathbf{z}),\\
~~~~\text{s.t.} ~~&\mathbb{E}[|\mathbf{x}|^2] = P_x, \label{power_constrain} \\
&\int_{-\infty}^{\infty}p(x) dx = 1,\\
&\mathcal{D}(p_{_1}||p_{_0}) \leq 2\epsilon^2,\label{covert_constraint_10}\\
&p(x) \geq 0, \label{positive_x_constraint_10}
\end{align}
\end{subequations}
where $P_x^{\ast} = P_x^{\epsilon}$
and {$P_x^{\epsilon}$ is the solution to
\begin{align}
\frac{1}{2}\left(\frac{P_x^{\epsilon}}{\sigma_w^2} + \log \frac{\sigma_w^2}{P_x^{\epsilon} + \sigma_w^2}\right) = 2 \epsilon^2.
\end{align}}
\end{theorem}

In \eqref{opt10}, we have $p_{_1}(y) = \int_{-\infty}^{\infty}g_{n_w}(y-x)p(x)dx$ and $p_{_0}(y) = \mathcal{N}(0,\sigma_w^2)$, where $g_{n_w}(\cdot)$ denotes the probability density function (pdf) of $n_w$. We note that $p(x) = \mathcal{N}(0,P_x)$ maximizes $I(\mathbf{x},\mathbf{z})$ subject to $\mathbb{E}[|x|^2] = P_x$ \cite[Theorem 8.6.5]{cover2002elements} and the maximum $I(\mathbf{x},\mathbf{z})$ is a monotonically increasing function of $P_x$. As such, we can prove Theorem~1 in two steps. In the first step, we prove that $p(x) = \mathcal{N}(0,P_x)$ minimizes $\mathcal{D}(p_{_1}||p_{_0})$ subject to $\mathbb{E}[|\mathbf{x}|^2] = P_x$ and $\int_{-\infty}^{\infty}p(x) dx = 1$, which is detailed in the following Section~III-A. In the second step, we determine the optimal value of $P_x$ that maximizes $I(\mathbf{x},\mathbf{z})$ subject to $\mathbb{E}[|\mathbf{x}|^2] = P_x$, $\int_{-\infty}^{\infty}p(x) dx = 1$, and $\mathcal{D}(p_{_1}||p_{_0}) \leq 2\epsilon^2$, which is presented in Section~III-B.

\subsection{Zero-Mean Gaussian Signalling is Optimal}

In this subsection, we present the first step in the proof of Theorem~\ref{theorem1}. Specifically, we prove the following theorem.

\begin{theorem}\label{theorem2}
The zero-mean Gaussian distributed $\mathbf{y}$ with variance $P_y$, i.e., $p_{_1}(y) = \mathcal{N}(0,P_y)$, is the solution to the following optimization problem
\begin{subequations}\label{opt10_sub}
\begin{align}
 \argmin_{p_{_1}(y)} ~~&\mathcal{D}(p_{_1}||p_{_0}),\\
~~~~\text{s.t.} ~~&\mathbb{E}[|\mathbf{y}|^2] = \int_{-\infty}^{\infty} y^2 p_{_1}(y) dy = P_y, \label{power_constraint_10} \\
&\int_{-\infty}^{\infty}p_{_1}(y) dy = 1, \label{pdf_constraint_10}\\
&p_{_1}(y) \geq 0, \label{positive_constraint_10}
\end{align}
\end{subequations}
where $P_y = P_x + \sigma_w^2$.
\end{theorem}
\begin{IEEEproof}
In order to prove Theorem~\ref{theorem2}, we first identify the solution of $p_{_1}(y)$ that minimizes $\mathcal{D}(p_{_1}||p_{_0})$ subject to \eqref{power_constraint_10} and \eqref{pdf_constraint_10} by using calculus of variations, and then prove that this solution also satisfies the constraint \eqref{positive_constraint_10}.
Following \eqref{opt10_sub}, we can write the functional of minimizing $\mathcal{D}(p_{_1}||p_{_0})$ subject to \eqref{power_constraint_10} and \eqref{pdf_constraint_10} as
\begin{align}\label{functional_orignal_10}
&\mathcal{D}(p_{_1}||p_{_0}) + \rho_0 \left[\int_{-\infty}^{\infty} p_{_1}(y) dy -1\right] \notag\\
&+ \rho_1 \left[\int_{-\infty}^{\infty} y^2 p_{_1}(y) dy - P_y\right] = \int_{-\infty}^{\infty} \mathcal{L}(y, p_{_1}(y)) dy - \tau,
\end{align}
where $\rho_0$ and $\rho_1$ are the Lagrange multipliers, which can be determined by the associated constraints later. Following \eqref{KL_10} and \eqref{functional_orignal_10}, $\mathcal{L} (y, p_{_1}(y))$ is given by
\begin{align}\label{functional_10}
\mathcal{L} (y, p_{_1}(y))  \!=\! p_{_1}(y) \log\frac{p_{_1}(y)}{p_{_0}(y)} \!+\! \rho_0 p_{_1}(y) \!+\! \rho_1 y^2 p_{_1}(y),
\end{align}
and $\tau$ is a constant given by
\begin{align}
\tau = \rho_0 + \rho_1 P_y.
\end{align}
Following \eqref{functional_10}, the functional derivative (i.e., the first derivative of $\mathcal{L} (y, p_{_1}(y))$ with respect to $p_{_1}(y)$) is given by
\begin{align}\label{first_derivation}
\frac{\partial \mathcal{L} (y, p_{_1}(y))}{\partial p_{_1}(y)} = \log \frac{p_{_1}(y)}{p_{_0}(y)} + 1 + \rho_0 + \rho_1 y^2.
\end{align}
Using the calculus of variations, a necessary condition on the solution to minimizing $\mathcal{D}(p_{_1}||p_{_0})$ subject to \eqref{power_constraint_10} and \eqref{pdf_constraint_10} is that this solution guarantees the functional derivative given in \eqref{first_derivation} being zero \cite{gelfand2000calculus}. As such, setting ${\partial \mathcal{L} (y, p_{_1}(y))}/{\partial p_{_1}(y)}= 0$, we have the solution given by
\begin{align}\label{p1_format_10}
p_{_1}(y) = p_{_0}(y)e^{-\rho_1 y^2 - \rho_0 - 1}.
\end{align}

We next determine the values of $\rho_0$ and $\rho_1$ based on the constraints given in \eqref{power_constraint_10} and \eqref{pdf_constraint_10}. Substituting $p_{_0}(y) = \mathcal{N}(0, \sigma_w^2)$ into \eqref{p1_format_10}, we have
\begin{align}\label{pdf_derive_2}
\int_{-\infty}^{\infty}p_{_1}(y) dy &= \frac{2}{\sqrt{2 \pi}\sigma_w}e^{\rho_0 +1}\int_{0}^{\infty}e^{-\left(\frac{1}{2 \sigma_w^2} +\rho_1\right) y^2} dy \notag\\
&=\frac{e^{-\rho_0 -1}}{\sqrt{1 + 2\rho_1 \sigma_w^2}},
\end{align}
{where the identity \cite[Eq. (3.321.3)]{gradshteuin2007table}
\begin{align}
\int_0^{\infty} e^{-q^2 x^2} dx = \frac{\sqrt{\pi}}{2 q}
\end{align}}
is applied to compute the integral in \eqref{pdf_derive_2}. We note that ${1}/{2 \sigma_w^2} -\rho_1 > 0$ is required in \eqref{pdf_derive_2} for optimality and from \eqref{pdf_constraint_10} we have
\begin{align}\label{pdf_derive_10}
{e^{-\rho_0 -1}}={\sqrt{1 + 2\rho_1 \sigma_w^2}}.
\end{align}
Again, substituting $p_{_0}(y) = \mathcal{N}(0, \sigma_w^2)$ into \eqref{p1_format_10}, we have
\begin{align}\label{power_derive_10}
\int_{-\infty}^{\infty}y^2 p_{_1}(y) dy &= \frac{2}{\sqrt{2 \pi}\sigma_w}e^{-\rho_0 -1}\int_{0}^{\infty}y^2 e^{-\left(\frac{1}{2 \sigma_w^2} +\rho_1\right) y^2} dy \notag\\
&=\frac{e^{-\rho_0 -1} \sigma_w^2}{(1 + 2\rho_1 \sigma_w^2)^{3/2}},
\end{align}
{where the identity \cite[Eq. (3.326.2)]{gradshteuin2007table}
\begin{align}
\int_0^{\infty} x^2 e^{-q^2 x^2} dx = \frac{\sqrt{\pi}}{4 q^3}
\end{align}}
is applied to compute the integral in \eqref{power_derive_10}.
Following \eqref{power_constraint_10} and substituting \eqref{pdf_derive_10} into \eqref{power_derive_10}, we have
\begin{align}\label{rho1_expression}
\rho_1 = -\frac{1}{2\sigma_w^2} + \frac{1}{2 P_y}.
\end{align}
We note that the value of $\rho_1$ given in \eqref{rho1_expression} guarantees ${1}/{2 \sigma_w^2} -\rho_1 > 0$. Finally, substituting \eqref{pdf_derive_10} and \eqref{rho1_expression} into \eqref{p1_format_10}, we have
\begin{align}\label{p1y_sol}
p_{_1}(y) = \frac{1}{\sqrt{2 \pi P_y}} e^{-\frac{y^2}{2 P_y}},
\end{align}
which indicates that $p_{_1}(y)$ is a Gaussian distribution with zero mean and variance $P_y$.

We next prove that $p_{_1}(y)$ given in \eqref{p1y_sol} satisfies a sufficient condition to be a solution to minimizing $\mathcal{D}(p_{_1}||p_{_0})$ subject to \eqref{power_constraint_10} and \eqref{pdf_constraint_10}.
To this end, following \eqref{first_derivation} the second derivative of $\mathcal{L} (y, p_{_1}(y))$ with respect to $p_{_1}(y)$ is derived as
\begin{align}\label{second_derivation}
\frac{\partial \mathcal{L}^2 (y, p_{_1}(y))}{\partial p_{_1}^2(y)} = \frac{1}{p_{_1}(y)}.
\end{align}
From \eqref{second_derivation}, we have $\frac{\partial \mathcal{L}^2 (y, p_{_1}(y))}{\partial p_{_1}^2(y)} \geq k \|p_{_1}(y)\|^2$ for all $y$ and for some constant $k > 0$ (which is the sufficient condition for $p_{_1}(y)$ being the solution \cite{gelfand2000calculus}), since as per \eqref{p1y_sol} we have $0 \leq p_{_1}(y) \leq {1}/{\sqrt{2 \pi P_y}}$. Specifically, in order to guarantee ${1}/{p_{_1}(y)} \geq k \|p_{_1}(y)\|^2$ for all $y$ we can select any value of $k$ within $0 < k \leq 2 \pi P_y \sqrt{2 \pi P_y}$.
So far, we have proved that $p_{_1}(y)$ given in \eqref{p1y_sol} is the solution to minimizing $\mathcal{D}(p_{_1}||p_{_0})$ subject to \eqref{power_constraint_10} and \eqref{pdf_constraint_10}, and clearly it also satisfies \eqref{positive_constraint_10}. We conclude that $p_{_1}(y)$ given in \eqref{p1y_sol} is the solution to the optimization problems given in \eqref{opt10_sub}. This completes the proof of Theorem~\ref{theorem2}.
%
\end{IEEEproof}

Following \eqref{hypothses} and noting $\mathbf{n}_w \sim \mathcal{N}(0,\sigma_w^2)$, Theorem~\ref{theorem2} indicates that the optimal distribution of $\mathbf{x}$ that minimizes $\mathcal{D}(p_{_1}||p_{_0})$ is a Gaussian distribution with zero mean. Together with \cite[Theorem 8.6.5]{cover2002elements}, we can conclude that the solution to the optimization problem given in \eqref{opt10} is that $\mathbf{x}$ follows a Gaussian distribution with zero mean. We next determine the variance of this zero-mean Gaussian distributed $\mathbf{x}$ in the following subsection.

\subsection{Optimal Transmit Power}

In this subsection, we present the second step in the proof of Theorem~\ref{theorem1}. Specifically, we derive the optimal value of $P_x$, i.e., the variance of $\mathbf{x}$ with a zero-mean Gaussian distribution. To this end, we first prove that $\mathcal{D}(p_{_1}||p_{_0})$ is a monotonically increasing function of $P_y$ and thus of $P_x$ for $p_{_1}(y)$ given in \eqref{p1y_sol}.

Substituting $p_{_0}(y) = \mathcal{N}(0, \sigma_w^2)$ and \eqref{p1y_sol} into \eqref{KL_10}, we have
\begin{align}\label{KL_10_gaussian}
\mathcal{D}(p_{_1}||p_{_0}) = \frac{1}{2}\left(\frac{P_y}{\sigma_w^2} - 1 + \log \frac{\sigma_w^2}{P_y}\right).
\end{align}
Then, the first derivative of $\mathcal{D}(p_{_1}||p_{_0})$ with respect to $P_y$ is given by
\begin{align}\label{KL_10_gaussian_der}
\frac{\partial \mathcal{D}(p_{_1}||p_{_0})}{\partial P_y} = \frac{1}{2}\left(\frac{1}{\sigma_w^2} - \frac{1}{P_y}\right),
\end{align}
which is non-negative since $P_y = P_x + \sigma_w^2 > \sigma_w^2$. This indicates that $\mathcal{D}(p_{_1}||p_{_0})$ monotonically increases with $P_y$ and thus with $P_x$. We denote the value of $P_x$ that guarantees $\mathcal{D}(p_{_1}||p_{_0}) = 2 \epsilon^2$ as $P_x^{\epsilon}$. Following \eqref{KL_10_gaussian}, $P_x^{\epsilon}$ is the value of $P_x$ that guarantees
\begin{align}
\frac{1}{2}\left(\frac{P_x + \sigma_w^2}{\sigma_w^2} - 1 + \log \frac{\sigma_w^2}{P_x + \sigma_w^2}\right) = 2 \epsilon^2.
\end{align}
Noting the fact that the maximum $I(\mathbf{x},\mathbf{z})$ achieved by $p(x) = \mathcal{N}(0,P_x)$ without the covertness constraint also monotonically increases with $P_x$ as per \cite[Theorem 8.6.5]{cover2002elements}, we can conclude $P_x^{\ast} = P_x^{\epsilon}$. This completes the proof of Theorem~\ref{theorem1}.

Theorem~\ref{theorem1} indicates that Gaussian signalling can simultaneously achieve the maximum mutual information from Alice to Bob and ensure a minimum KL divergence from the likelihood function under $\Hnull$ to that under $\Halt$ at Willie. As such, it is the optimal signalling for covert communications with $\mathcal{D}(p_{_1}||p_{_0}) \leq 2\epsilon^2$ as the covertness constraint.

\section{With $\mathcal{D}(p_{_0}||p_{_1}) \leq 2\epsilon^2$ as the Covertness Constraint}\label{section_01}

{In this section, we analytically prove that Gaussian signaling is \textbf{not} optimal for the covert communication with $\mathcal{D}(p_{_0}||p_{_1}) \leq 2\epsilon^2$ as the constraint. We also present a skew-normal signalling strategy as a benchmark and derive the expression of $p_{_1}(y)$ for this skew-normal signalling in this section, which allows us to numerically show that skew-normal signalling can be better than Gaussian signalling when  $\mathcal{D}(p_{_0}||p_{_1}) \leq 2\epsilon^2$ is used as the covertness constraint in our numerical results (i.e., Section VI).}

\subsection{Gaussian Signalling is Not Optimal}

In this subsection, we prove that Gaussian signaling is \textbf{not} optimal for covert communication with $\mathcal{D}(p_{_0}||p_{_1}) \leq 2\epsilon^2$ as the constraint in the following theorem.

\begin{theorem}\label{theorem3}
Gaussian signaling, i.e., $p(x) = \mathcal{N}(m_x, \sigma_x^2)$, is not the solution to the following optimization problem
\begin{subequations}\label{opt01}
\begin{align}
 \argmax_{p(x), P_x} ~~&I(\mathbf{x},\mathbf{z}),\\
~~~~\text{s.t.} ~~&\mathbb{E}[|\mathbf{x}|^2] = P_x, \label{power_constrain01} \\
&\int_{-\infty}^{\infty}p(x) dx = 1,\label{pdf_constraint_01}\\
&\mathcal{D}(p_{_0}||p_{_1}) \leq 2\epsilon^2, \label{covert_constraint_01}\\
&p(x) \geq 0, \label{positive_x_constraint_01}
\end{align}
\end{subequations}
where $m_x$ and $\sigma_x^2$ can take arbitrary values.
\end{theorem}
\begin{IEEEproof}
In order to prove Theorem~\ref{theorem3}, we next prove that Gaussian signalling is not in general the solution to the optimization problem given \eqref{opt01} in a special case, where Bob and Willie both experience the same level of AWGN. In this special case, we have $\mathbf{n}_w$ in \eqref{hypothses} and $\mathbf{n}_b$ in \eqref{xn} are i.i.d. and thus the pdf of $\mathbf{z}$ and the pdf of $\mathbf{y}$ under $\Halt$ are the same, i.e., we have $p(z) = p_{_1}(y)$. As such, in the rest of this proof we use $p_{_1}(y)$ to represent $p(z)$.
Following \eqref{xn} and noting that $\mathbf{x}$ is independent of $\mathbf{n}_b$, we have
\begin{align}\label{mutual_inf}
I(\mathbf{x},\mathbf{z}) = h(\mathbf{z}) - h(\mathbf{n}_b),
\end{align}
where
\begin{align}
h(z) = \int_{-\infty}^{\infty} p(z) \log \frac{1}{p(z)} dz = \int_{-\infty}^{\infty} p_{_1}(y) \log \frac{1}{p_{_1}(y)} dy
\end{align}
is the differential entropy of $\mathbf{z}$ and $h(\mathbf{n}_b)$ is the differential entropy of $\mathbf{n}_b$, which is not a function of $p(z)$ or $p_{_1}(y)$. As such, in this special case to prove Theorem~\ref{theorem3} we are going to prove that $p_{_1}(y) = \mathcal{N}(0, P_y)$ is not the solution to the following optimization problem:
\begin{subequations}\label{opt01_special}
\begin{align}
 \argmax_{p_{_1}(y)} ~~& \int_{-\infty}^{\infty} p_{_1}(y) \log \frac{1}{p_{_1}(y)} dy,\\
~~~~\text{s.t.} ~~&\int_{-\infty}^{\infty} p_{_1}(y) dy = 1, \label{pdf_constraint_01_special} \\
&\int_{-\infty}^{\infty}y^2 p_{_1}(y) dy = P_y, \label{power_constrain01_special}\\
&\mathcal{D}(p_{_0}||p_{_1}) \leq 2 \epsilon^2, \label{covert_constraint_01_special}\\
& p_{_1}(y) \geq 0. \label{positive_x_constraint_01_special}
\end{align}
\end{subequations}
In order to apply calculus of variations, following \eqref{opt01_special} we can write the functional as
\begin{align}\label{functional_ori_01}
&\int_{-\infty}^{\infty} p_{_1}(y) \log \frac{1}{p_{_1}(y)} dy + \eta_0 \left[\mathcal{D}(p_{_0}||p_{_1})-2\epsilon^2\right] \notag\\
&+ \eta_1 \left[\int_{-\infty}^{\infty} p_{_1}(y) dy -1\right] + \eta_2 \left[\int_{-\infty}^{\infty} y^2 p_{_1}(y) dy - P_y\right]\notag \\
& = \int_{-\infty}^{\infty} \overline{\mathcal{L}}(y, p_{_1}(y)) dy - c,
\end{align}
where $\eta_0$, $\eta_1$, $\eta_2$, and $\eta_3$ are the Lagrange multipliers.
Then, $\overline{\mathcal{L}}(y, p_{_1}(y))$ in \eqref{functional_ori_01} is given by
\begin{align}\label{functional_01}
\overline{\mathcal{L}}(y, p_{_1}(y))  &= p_{_1}(y) \log \frac{1}{p_{_1}(y)} + \eta_0 p_{_0}(y) \log\frac{p_{_0}(y)}{p_{_1}(y)}\notag \\
&~~~~+ \eta_1 p_{_1}(y) + \eta_2 y^2 p_{_1}(y),
\end{align}
and $c$ is a constant determined by the Lagrange multipliers, $h(\mathbf{n}_b)$, $\epsilon^2$, and $P_y$.
Following \eqref{functional_01}, the functional derivative (i.e., the first derivative of $\overline{\mathcal{L}}(y, p_{_1}(y))$ with respect to $p_{_1}(y)$) is given by
\begin{align}\label{functional_der_01}
\frac{\partial \overline{\mathcal{L}}(y, p_{_1}(y))}{\partial p_{_1}(y)} \!=\! \!-\!\log p_{_1}(y) \!-\!1 \!-\! \eta_0 \frac{p_{_0}(y)}{p_{_1}(y)} \!+\! \eta_1 \!+\! \eta_2 y^2.
\end{align}
Using the calculus of variations, a necessary condition for the optimal $p_{_1}(y)$ in \eqref{opt01_special} is the existence of Lagrange multipliers such that the functional derivative given in \eqref{functional_der_01} is zero. As per \cite[Theorem 8.6.5]{cover2002elements}, $p_{_1}(y) = \mathcal(N)(0, P_y)$ maximizes the mutual information between $x$ and $z$ subject to the constraints given in \eqref{pdf_constraint_01_special}, \eqref{power_constrain01_special}, and \eqref{positive_x_constraint_01_special}. As such, $p_{_1}(y) = \mathcal{N}(0, P_y)$ must satisfy
\begin{align}\label{functional_der_01_p1}
-\log p_{_1}(y) -1 + \eta_1^a + \eta_2^a y^2 = 0,
\end{align}
for two Lagrange multipliers $\eta_1^a$ and $\eta_2^a$. If $p_{_1}(y) = \mathcal{N}(0, P_y)$ is the solution to the optimization problem given in \eqref{opt01_special}, following \eqref{functional_der_01_p1} it must satisfy
\begin{align}\label{functional_der_01_p2}
- \eta_0 \frac{p_{_0}(y)}{p_{_1}(y)} + \eta_1^b + \eta_2^b y^2 = 0,
\end{align}
with $\eta_1^b = \eta_1 - \eta_1^a$ and $\eta_2^b = \eta_2 - \eta_2^a$. If \eqref{functional_der_01_p2} is satisfied, then $p_{_1}(y)$ is given by
\begin{align}\label{p1y_format_non}
p_{_1}(y) = \frac{\eta_0 p_{_0}(y)}{\eta_1^b + \eta_2^b y^2}.
\end{align}
We note that in \eqref{p1y_format_non} the value of $\eta_2^b$ cannot be zero. Otherwise, we will have $p_{_1}(y) = \eta_0 p_{_0}(y)/\eta_1^b$. In order to guarantee the pdf constraint \eqref{pdf_constraint_01} with $p_{_1}(y) = \eta_0 p_{_0}(y)/\eta_1^b$, we would have $\eta_0/\eta_1^b = 1$, which cannot guarantee the power constraint \eqref{power_constrain01} simultaneously, since $P_y = P_x + \sigma_w^2 > \sigma_w^2$. As such, \eqref{p1y_format_non} with $\eta_2^b \neq 0$ indicates that the optimal signalling (if it exists) is not Gaussian, which completes the proof of Theorem~\ref{theorem3}.
\end{IEEEproof}

\subsection{A Benchmark $p(x)$: Skew-Normal Distribution}

{In this subsection, we present the skew-normal distribution as a benchmark $p(x)$, where we consider the case in which the AWGN at Bob and Willie is i.i.d (i.e., $\mathbf{n}_w$ and $\mathbf{n}_b$ are i.i.d) such that the received signal at Willie $\mathbf{y}$ and the received signal at Bob $\mathbf{z}$ follow the same distribution. We derive the expression of $p_{_1}(y)$ for this skew-normal distribution, which allows us to numerically show that it can be better than Gaussian signalling when the covertness constraint is given by $\mathcal{D}(p_{_0}||p_{_1}) \leq 2\epsilon^2$.}

If $\mathbf{x}$ follows a skew-normal distribution, the corresponding expression of $p(x)$ is given by \cite{azzalini2014the}
\begin{align}\label{pdf_skew_normal}
p(x) = \frac{1}{\omega \sqrt{2 \pi}} e^{-\frac{(x-\mu)^2}{2 \omega^2}}\left[1 + \text{erf}\left(\frac{\theta (x - \mu)}{\omega \sqrt{2}}\right)\right],
\end{align}
where $\mu$ is the location parameter, $\omega$ is the scale parameter, $\theta$ is the skew parameter, and $\text{erf}(x)$ is the error function given by $\text{erf}(x) = \frac{1}{\sqrt{\pi}}\int^x_{-x}e^{-t^2}dt$. We note that the normal distribution is recovered from \eqref{pdf_skew_normal} when $\theta =0$ and the skewness increases as $|\theta|$  increases. In addition, the skew-normal distribution is right skewed relative to the normal distribution if $\theta >0$ and is left skewed if $\theta <0$. For the distribution given in \eqref{pdf_skew_normal}, the mean and variance of $x$ are, respectively, given by
\begin{align}
\mathbb{E}[\mathbf{x}] &= \mu + \omega \delta \sqrt{\frac{2}{\pi}},\label{skew_mean}\\
\mathbb{E}[|\mathbf{x} - \mathbb{E}[\mathbf{x}]|^2] &= \omega^2 \left(1-\frac{2 \delta^2}{\pi}\right),\label{skew_variance}
\end{align}
where $\delta = \theta/\sqrt{1 + \theta^2}$. In this work, we focus on the skew-normal distribution with zero and $P_x$ as the mean and variance, respectively. To this end, as per \eqref{skew_mean} and \eqref{skew_variance}, for a given $\theta$ we have
\begin{align}
\omega &= \pm \sqrt{\frac{P_x}{1 - \frac{2 \theta^2}{\pi (1 + \theta^2)}}},\label{omega}\\
\mu &= -\omega \sqrt{\frac{2 \theta^2}{\pi (1 + \theta^2)}}. \label{mu}
\end{align}
We can vary the values of $\theta$ to obtain different skew-normal distributions with zero and $P_x$ as the mean and variance, respectively, where the values of $\omega$ and $\mu$ are updated as per $\theta$ according to \eqref{omega} and \eqref{mu}, respectively. This allows us to find a potential better $p(x)$ than the normal distribution in terms of achieving a higher $I(\mathbf{x},\mathbf{z})$ subject to the constraints given in \eqref{power_constrain01}, \eqref{pdf_constraint_01}, \eqref{covert_constraint_01}, and \eqref{positive_x_constraint_01}, which will be confirmed in the numerical section (i.e., Section VI).

In order to facilitate the calculation of the KL divergence from $p_{_0}(y)$ to $p_{_1}(y)$ and the mutual information between $\mathbf{x}$ and $\mathbf{z}$, we derive the expression of $p_{_1}(y)$ for the skew-normal $p(x)$ in the following proposition, which is also the expression of $p(z)$ for i.i.d. $\mathbf{n}_w$ and $\mathbf{n}_b$.

\begin{proposition}\label{proposition_add}
For a skew-normal $p(x)$ with zero mean, variance $P_x$, and a non-zero skew parameter $\theta$, following \eqref{hypothses} the expression of $p_{_1}(y)$ is derived as
\begin{align}\label{sum_skew_normal_pdf}
&p_{_1}(y) = \frac{|\omega|}{\omega \sqrt{2 \pi (\sigma_w^2 + \omega^2)}}e^{-\frac{(y - \mu)^2}{2(\sigma_w^2 + \omega^2)}}\notag \\
&+
\frac{1}{\pi\sqrt{2\pi}\sigma_w^3 \theta^2}\sum_{k = 1}^{\infty} \frac{(-1)^{k+1}e^{-\frac{(y-\mu)^2}{2 \sigma_w^2}}}{(2k-1)(k-1)!}\left(\frac{(\sigma_w^2 + \omega)^2}{\sigma_w^2 \theta^2}\right)^{-\frac{1}{2}-k}\notag \\
&\times \left[-\frac{\sigma_w \theta}{|\theta|}(\sigma_w^2 + \omega^2)^{k+1} \Gamma(k)~\!\!{_1}F{_1}\!\!\left(k, \frac{1}{2}, \frac{\omega^2 (y-\mu)^2}{2 \sigma_w^2 (\sigma_w^2 + \omega^2)}\right)\right. \notag \\
&\left.+ \frac{\sigma_w^{2k+3}\theta^3}{|\theta|^{-2k+1}}\left(\frac{\sigma_w^2 + \omega^2}{\sigma^2 \theta^2}\right)^{k+1}\Gamma(k)~\!\!{_1}F{_1}\!\!\left(k, \frac{1}{2}, \frac{\omega^2 (y-\mu)^2}{2 \sigma_w^2 (\sigma_w^2 + \omega^2)}\right)\right.\notag \\
&\left.+ 2\sqrt{2} \omega (\sigma_w^2 + \omega^2)^{k + \frac{1}{2}}(y - \mu) \Gamma\left(k+\frac{1}{2}\right)\right.\notag\\
&~~~~\times
\left.\!\!{_1}F{_1}\!\!\left(k+\frac{1}{2}, \frac{3}{2}, \frac{\omega^2 (y-\mu)^2}{2 \sigma_w^2 (\sigma_w^2 + \omega^2)}\right)\right],
\end{align}
where ${_1}F{_1}(a,b,z)$ is the Kummer confluent hypergeometric function.
\end{proposition}
\begin{IEEEproof}
Following \eqref{hypothses}, we have $\mathbf{y} = \mathbf{x} + \mathbf{n}_w$ under $\Halt$ and noting $\mathbf{n}_w \sim \mathcal{N}(0,\sigma_w^2)$ we have
\begin{align}\label{skew_normal_proof1}
p_{_1}(y) = \frac{1} {\sqrt{2 \pi} \sigma_w} \int_{-\infty}^{\infty} e^{-\frac{(y-x)^2}{2 \sigma_w^2}}p(x) dx,
\end{align}
since $\mathbf{x}$ and $\mathbf{n}_w$ are independent. Then, substituting \eqref{pdf_skew_normal} into \eqref{skew_normal_proof1} we have
\begin{align}\label{skew_normal_proof2}
&p_{_1}(y) = \frac{1} {{2 \pi} \sigma_w \omega} \int_{-\infty}^{\infty} e^{-\frac{(y-x)^2}{2 \sigma_w^2}-\frac{(x-\mu)^2}{2 \omega^2}} dx \notag \\
&~~~~~~~~~~+ \frac{1} {{2 \pi} \sigma_w \omega} \int_{-\infty}^{\infty} e^{-\frac{(y-x)^2}{2 \sigma_w^2}} \text{erf}\left(\frac{\theta (x - \mu)}{\omega \sqrt{2}}\right) dx \notag \\
&\overset{a}{=} \frac{|\omega|}{\omega \sqrt{2 \pi (\sigma_w^2 + \omega^2)}}e^{-\frac{(y - \mu)^2}{2(\sigma_w^2 + \omega^2)}}\notag \\
& \!+\! \frac{\sqrt{2}} {{ \pi \sqrt{\pi}} \sigma_w \theta} \sum_{k = 1}^{\infty} \frac{(\!-\!1)^{k+1}}{(2k\!-\!1)(k\!-\!1)!}\int_{\!-\!\infty}^{\infty} \chi^{2k-1} e^{\!-\!\frac{(y \!-\! \frac{\omega \sqrt{2} \chi}{\theta}-\mu)^2}{2 \sigma_w^2}} d \chi,
\end{align}
where $\overset{a}{=}$ is achieved by setting $\chi = \frac{\theta (x - \mu)}{\omega \sqrt{2}}$ and with the aid of the following identity \cite[Eq. (8.253.1)]{gradshteuin2007table}
\begin{align}
\text{erf}(\chi) &= \frac{2}{\sqrt{\pi}}\sum_{k = 1}^{\infty} (-1)^{k+1}\frac{\chi^{2k-1}}{(2k-1)(k-1)!}.
\end{align}
Then, solving the resultant integrals in \eqref{skew_normal_proof2} leads to the desired result in \eqref{sum_skew_normal_pdf}, which completes the proof of Proposition~\ref{proposition_add}.
\end{IEEEproof}

Following Proposition~\ref{proposition_add}, the KL divergence from $p_{_0}(y)$ to $p_{_1}(y)$ can be obtained by substituting \eqref{sum_skew_normal_pdf} into \eqref{KL_qp}. Since $\mathbf{x}$ and $\mathbf{n}_b$ are i.i.d, the mutual information between $\mathbf{x}$ and $\mathbf{z}$ can be written as
\begin{align}
I(\mathbf{x},\mathbf{z}) &= h(\mathbf{z}) - h(\mathbf{n}_b)\notag \\
&= -\int_{-\infty}^{\infty} p(z) \log p(z) dz - \frac{1}{2}\log(2 \pi e \sigma_b^2),
\end{align}
where the expression for $p(z)$ is the same as that for $p_{_1}(y)$ given in \eqref{sum_skew_normal_pdf}.

\section{Covert Communications with Gaussian Signalling}\label{section_gaussian}

In this section, we first present Willie's detection performance in terms of the minimum detection error probability (i.e., $\xi^{\ast}$) with Gaussian signalling ($x$ follows the zero-mean Gaussian distribution with variance $P_x$, i.e., $p(x) = \mathcal{N}(0, P_x)$). Then, we examine the tightness of the two lower bounds on $\xi^{\ast}$ determined by the two KL divergences, i.e., $\mathcal{D}(p_{_0}||p_{_1})$ and $\mathcal{D}(p_{_1}||p_{_0})$, based on which we conclude that $\mathcal{D}(p_{_0}||p_{_1}) \leq 2 \epsilon^2$ is a more desirable covertness constraint than $\mathcal{D}(p_{_1}||p_{_0}) \leq 2 \epsilon^2$.

\subsection{Willie's Detection Performance}

With $p(x) = \mathcal{N}(0, P_x)$, as per \eqref{hypothses} the likelihood functions of $\mathbf{y}$ under $\Hnull$ and $\Halt$ are given as
\begin{align}
p_{_0}(y) &= \mathcal{N}(0, P_x),\label{likeli_H0}\\
p_{_1}(y) &= \mathcal{N}(0, P_x + \sigma_w^2), \label{likeli_H1}
\end{align}
respectively. Considering the equal \emph{a priori} probabilities for $\Hnull$ and $\Halt$, the optimal test that minimizes $\xi$ is the likelihood ratio test with $1$ as the optimal detection threshold \cite{bernard2008principles,yan2016location}, which is given by
\begin{equation}\label{LRT}
\frac{p_{_1}(y)}{p_{_0}(y)} \begin{array}{c}
\overset{\Hoalt}{\geq} \\
\underset{\Honull}{<}
\end{array}%
1.
\end{equation}
After some algebraic manipulations, \eqref{LRT} can be reformulated as
\begin{equation}\label{rodiometer}
y^2 \begin{array}{c}
\overset{\Hoalt}{\geq} \\
\underset{\Honull}{<}
\end{array}%
\phi^{\ast},
\end{equation}
where $\phi^{\ast}$ is the optimal threshold for $y^2$, which is given by
\begin{align}\label{optimal_t_FTP}
\phi^{\ast} = \frac{(P_x + \sigma_w^2) \sigma_w^2}{P_x} \ln \left(\frac{P_x + \sigma_w^2}{\sigma_w^2}\right).
\end{align}
Following \eqref{likeli_H0} and \eqref{likeli_H1}, we have the cumulative density functions (cdfs) of $y^2$ under $\Hnull$ and $\Halt$ given by
\begin{align}
P_{_0}(y^2) &= \frac{1}{\Gamma(1/2)}\gamma\left(\frac{1}{2}, \frac{y^2}{2 \sigma_w^2}\right),\label{likeli_H0_y2}\\
P_{_1}(y^2) &= \frac{1}{\Gamma(1/2)}\gamma\left(\frac{1}{2}, \frac{y^2}{2(P_x+\sigma_w^2)}\right), \label{likeli_H1_y2}
\end{align}
respectively, where $\gamma(\cdot, \cdot)$ is the lower incomplete gamma function given by $\gamma(n, x) = \int_0^x e^{-t} t^{n-1} dt$.
Then, following \eqref{rodiometer} the false positive and miss detection rates for this optimal detection threshold $\phi^{\ast}$ are derived as
\begin{align}
\alpha^{\ast} &\!=\! \Pr(\mathbf{y}^2 > \phi^{\ast}|\Hnull) \!=\! 1-  \frac{1}{\Gamma(1/2)}\gamma\left(\frac{1}{2}, \frac{\phi^{\ast}}{2 \sigma_w^2}\right), \label{a_gaussian}\\
\beta^{\ast} &\!=\! \Pr(\mathbf{y}^2 < \phi^{\ast}|\Halt) \!=\! \frac{1}{\Gamma(1/2)}\gamma\left(\frac{1}{2}, \frac{\phi^{\ast}}{2(P_x+\sigma_w^2)}\right),\label{b_gaussian}
\end{align}
respectively. We are going to use the above expressions of $\alpha^{\ast}$ and $\beta^{\ast}$ to evaluate the detection performance of Willie, based on which we can determine the maximum $I(\mathbf{x},\mathbf{z})$ achieved subject to the  covertness constraint $\xi^{\ast} = \alpha^{\ast} + \beta^{\ast} \geq 1 - \epsilon$.

\subsection{Mutual Information with Gaussian Signalling}

For $p(x) = \mathcal{N}(0, P_x)$, the mutual information of $\mathbf{x}$ and $\mathbf{z}$ as a function of $P_x$ is given by
\begin{align}\label{covert_capacity}
R_{ab}= \log\left(1 + \frac{P_x}{\sigma_b^2}\right).
\end{align}
Then, considering the covertness constraint $\xi^{\ast} \geq 1 - \epsilon$, the optimization problem at Alice is given by
\begin{subequations}\label{max_capacity_gaussian}
\begin{align}
 \argmax_{P_x>0} ~~&R_{ab},\\
~~~~\text{s.t.} ~~&\xi^{\ast} \geq 1 - \epsilon. \label{actual_covert_constraint}
\end{align}
\end{subequations}
Due to the complicated expressions of $\alpha^{\ast}$ and $\beta^{\ast}$ given in \eqref{a_gaussian} and \eqref{b_gaussian}, the solution to the optimization problem given in \eqref{max_capacity_gaussian} is mathematically intractable and thus we have to numerically search for it. Based on the searched optimal $P_x$, we will compare the achieved mutual information of $x$ and $z$ subject to $\xi^{\ast} \geq 1 - \epsilon$ with those achieved subject to $\mathcal{D}(p_{_0}||p_{_1}) \leq 2 \epsilon^2$ and $\mathcal{D}(p_{_1}||p_{_0}) \leq 2 \epsilon^2$ in Section~VI. To this end, in the following subsection we examine the difference between $\mathcal{D}(p_{_0}||p_{_1})$ and $\mathcal{D}(p_{_1}||p_{_0})$ with Gaussian signalling.

\subsection{Difference between $\mathcal{D}(p_{_0}||p_{_1})$ and $\mathcal{D}(p_{_1}||p_{_0})$}

In this subsection, we analytically prove $\mathcal{D}(p_{_0}||p_{_1}) \leq \mathcal{D}(p_{_1}||p_{_0})$ for Gaussian signalling, which leads to the fact that $\mathcal{D}(p_{_0}||p_{_1})$ determines a tighter lower bound on Willie's actual minimum detection error probability $\xi^{\ast}$ than $\mathcal{D}(p_{_1}||p_{_0})$ and thus $\mathcal{D}(p_{_0}||p_{_1}) \leq 2 \epsilon^2$ is a more desirable constraint in the covert communications with Gaussian signalling.
\begin{proposition}\label{proposition2}
For the zero-mean Gaussian signalling with $P_x$ as the transmit power, i.e., $p(x) = \mathcal{N}(0, P_x)$, we have
\begin{align}
\mathcal{D}(p_{_0}||p_{_1}) \leq \mathcal{D}(p_{_1}||p_{_0}),
\end{align}
where we recall that $P_y = P_x + \sigma_w^2 >\sigma_w^2$.
\end{proposition}
\begin{IEEEproof}
For $p(x) = \mathcal{N}(0, P_x)$, following \eqref{hypothses} and \eqref{KL_qp} we have
\begin{align}\label{KL_01_gaussian}
\mathcal{D}(p_{_0}||p_{_1}) = \frac{1}{2}\left(\frac{\sigma_w^2}{P_y} - 1 + \log \frac{P_y}{\sigma_w^2}\right).
\end{align}
Then, following \eqref{KL_10_gaussian} and \eqref{KL_01_gaussian} we have the difference between $\mathcal{D}(p_{_1}||p_{_0})$ and $\mathcal{D}(p_{_0}||p_{_1})$ as a function of $P_x$ given by
\begin{align}\label{difference}
&f(P_x) \triangleq \mathcal{D}(p_{_1}||p_{_0}) - \mathcal{D}(p_{_0}||p_{_1})\notag \\
&= \frac{P_y^2 - \sigma_w^4}{\sigma_w^2 P_y} - \log \frac{P_y^2}{\sigma_w^4}\notag\\
&=\frac{P_x + \sigma_w^2}{\sigma_w^2} - \frac{\sigma_w^2}{P_x + \sigma_w^2} - \frac{1}{2} \log(P_x + \sigma_w^2) + \log \sigma_w.
\end{align}
Following \eqref{difference}, the first derivative of $f(P_x)$ with respect to $P_x$ is derived as
\begin{align}\label{different_der}
\frac{\partial f(P_x)}{P_x} &= \frac{1}{\sigma_w^2} + \frac{\sigma_w^2}{(P_x+\sigma_w^2)^2} - \frac{1}{2(P_x + \sigma_w^2)} \notag \\
&= \frac{2 P_x^2 + 3\sigma_w^2 P_x + 3 \sigma_w^4}{2 \sigma_w^2 (P_x + \sigma_w^2)^2}\notag\\
&= \frac{1}{2 \sigma_w^2 (P_x + \sigma_w^2)^2}\left[2\left(P_x + \frac{3}{4}\sigma_w^2\right)^2 + \frac{15}{8}\sigma_w^2\right]\notag \\
& \geq 0,
\end{align}
due to $\sigma_w^2 \geq 0$ and $P_x > 0$. Then, as per \eqref{difference} and \eqref{different_der} we can conclude $\mathcal{D}(p_{_0}||p_{_1}) \leq \mathcal{D}(p_{_1}||p_{_0})$, which completes the proof of Proposition~\ref{proposition2}.
\end{IEEEproof}

Following Proposition~\ref{proposition2}, we have the following corollary with regard to the solutions to the optimization problems given in \eqref{opt10} and \eqref{opt01}.
\begin{corollary}\label{corollary1}
The solution to the optimization problem given in \eqref{opt10} is feasible to the optimization problem given in \eqref{opt01}, which leads to the fact that the maximum mutual information $I(\mathbf{x},\mathbf{z})$ achieved subject to $\mathcal{D}(p_{_0}||p_{_1}) \leq 2 \epsilon^2$ is higher than that achieved subject to $\mathcal{D}(p_{_1}||p_{_0}) \leq 2 \epsilon^2$.
\end{corollary}
\begin{IEEEproof}
Noting $\mathcal{D}(p_{_0}||p_{_1}) \leq \mathcal{D}(p_{_1}||p_{_0})$ as proved in Proposition~\ref{proposition2}, we can conclude that Gaussian signalling, i.e., $p(x) = \mathcal{N}(0,P_x^{\ast})$, which is the solution to \eqref{opt10}, is feasible to \eqref{opt01}. Noting the fact that both $\mathcal{D}(p_{_0}||p_{_1})$ and $\mathcal{D}(p_{_1}||p_{_0})$ are increasing function of $P_x$ for Gaussian signalling, we can conclude that the constraint $\mathcal{D}(p_{_0}||p_{_1}) \leq 2 \epsilon^2$  determines a higher value of $P_x$ than the constraint $\mathcal{D}(p_{_1}||p_{_0}) \leq 2 \epsilon^2$, which completes the proof.
\end{IEEEproof}
The gap between the maximum mutual information achieved subject to $\mathcal{D}(p_{_0}||p_{_1}) \leq 2 \epsilon^2$ and $\mathcal{D}(p_{_1}||p_{_0}) \leq 2 \epsilon^2$ will be explicitly examined for Gaussian signalling in our numerical section (i.e., Section VI). Following \eqref{distance}, \eqref{pinsker1}, \eqref{pinsker2}, and Proposition~\ref{proposition2}, we have the following corollary.
\begin{corollary}\label{corollary2}
For Gaussian signalling, we have
\begin{align}
\xi^{\ast} \geq 1\!-\!\sqrt{\mathcal{D}(p_{_0}||p_{_1})/2} \geq 1\!-\!\sqrt{\mathcal{D}(p_{_1}||p_{_0})/2},
\end{align}
which means that $\mathcal{D}(p_{_0}||p_{_1})$ determines a tighter lower bound on $\xi^{\ast}$ than $\mathcal{D}(p_{_1}||p_{_0})$ for Gaussian signalling.
\end{corollary}

Corollary~\ref{corollary2} indicates that for Gaussian signalling $\mathcal{D}(p_{_0}||p_{_1}) \leq 2 \epsilon^2$ is a more desirable constraint than $\mathcal{D}(p_{_1}||p_{_0}) \leq 2 \epsilon^2$ in practical covert communications.

\section{Numerical Results}\label{numerical_re}

In this section, we first present the KL divergence $\mathcal{D}(p_{_0}||p_{_1})$ and mutual information $I(\mathbf{x},\mathbf{z})$ for skew-normal signalling, which as shown can achieve a higher $I(\mathbf{x},\mathbf{z})$ subject to $\mathcal{D}(p_{_0}||p_{_1}) \leq 2 \epsilon^2$ than Gaussian signalling. This confirms that Gaussian signalling is not optimal for covert communications with $\mathcal{D}(p_{_0}||p_{_1}) \leq 2 \epsilon^2$ as the constraint. We then use $\xi^{\ast} \geq 1 - \epsilon$ (i.e., $\mathcal{V}_T(p_{_0}, p_{_1}) \leq \epsilon$) as the covertness constraint and numerically show that a skew-normal $p(x)$ can achieve a higher mutual information $I(\mathbf{x},\mathbf{z})$ than the normal $p(x)$, which draws a more general conclusion that Gaussian signalling is not optimal for covert communications with $\xi^{\ast} \geq 1 - \epsilon$ as the constraint. Finally, we numerically and explicitly examine the differences between covert communications with the aforementioned three different constraints for Gaussian signalling.

\begin{figure}[!t]
    \begin{center}
   {\includegraphics[width=3.5in, height=2.9in]{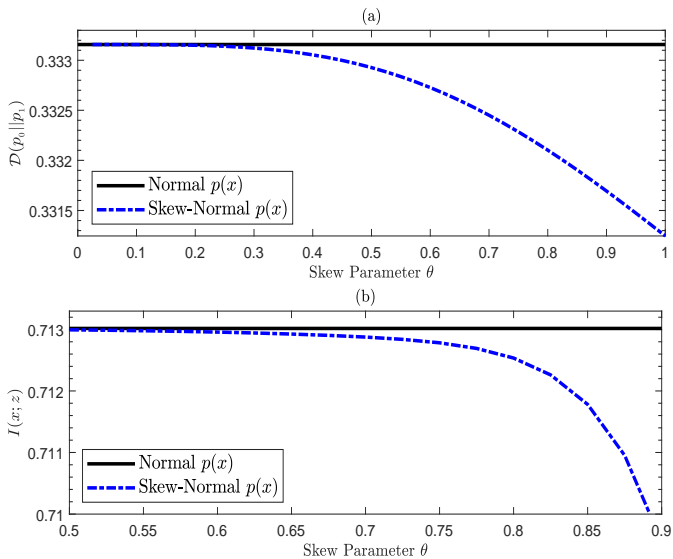}}
    \caption{The KL divergence $\mathcal{D}(p_{_0}||p_{_1})$ and mutual information $I(\mathbf{x},\mathbf{z})$ for skew-normal $p_{_1}(y)$ with different values of the skew parameter $\theta$, where $\sigma_b^2 = \sigma_w^2 = 0$dB and $P_x = 0$dB.}\label{fig:fig2}
    \end{center}
\end{figure}

\begin{figure}[!t]
    \begin{center}
   {\includegraphics[width=3.5in, height=2.9in]{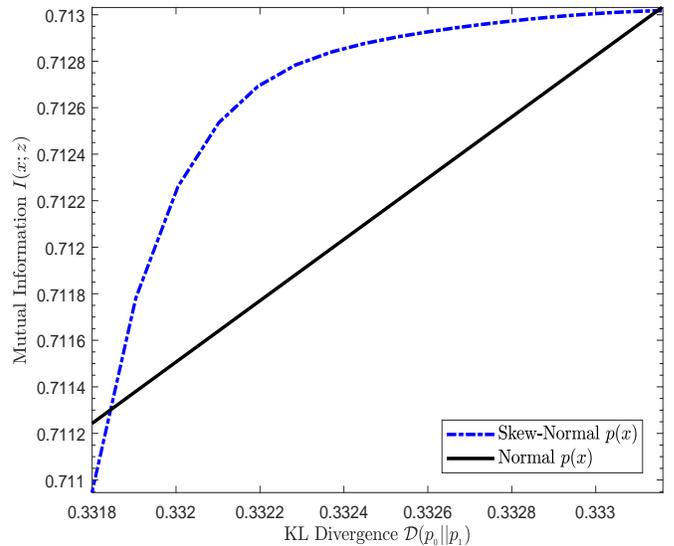}}
    \caption{The achieved mutual information $I(\mathbf{x},\mathbf{z})$ versus the associated KL divergence $\mathcal{D}(p_{_0}||p_{_1})$ for the skew-normal and normal $p(x)$.}\label{fig:fig3}
    \end{center}
\end{figure}

In Fig.~\ref{fig:fig2}, we plot the KL divergence $\mathcal{D}(p_{_0}||p_{_1})$ and mutual information $I(\mathbf{x},\mathbf{z})$ for a skew-normal $p(x)$ with different skew parameters, where the mean and variance of $x$ are fixed as $0$ and $P_x$, respectively. From this figure, we observe that the skew-normal $p(x)$ can achieve a lower KL divergence $\mathcal{D}(p_{_0}||p_{_1})$ with some specific values of the skew parameter $\theta$ than the corresponding normal $p(x)$, although the former always achieves a lower mutual information $I(\mathbf{x},\mathbf{z})$ than the later. This provides the possibility that the skew-normal $p(x)$ achieves a higher $I(\mathbf{x},\mathbf{z})$ subject to $\mathcal{D}(p_{_0}||p_{_1}) \leq 2 \epsilon^2$ than the normal $p(x)$. To confirm this, we plot the achieved mutual information $I(\mathbf{x},\mathbf{z})$ versus the associated KL divergence $\mathcal{D}(p_{_0}||p_{_1})$ for skew-normal and normal $p(x)$ in Fig.~\ref{fig:fig3}. In order to plot Fig.~\ref{fig:fig3}, we fix $P_x = 0$dB for the skew-normal $p(x)$ and vary $\theta$ to generate different values of $I(\mathbf{x},\mathbf{z})$ and $\mathcal{D}(p_{_0}||p_{_1})$, while for the normal $p(x)$ we slightly vary $P_x$ to obtain similar values of $I(\mathbf{x},\mathbf{z})$ and $\mathcal{D}(p_{_0}||p_{_1})$, since for the normal $p(x)$ there is a unique $I(\mathbf{x},\mathbf{z})$ and a unique $\mathcal{D}(p_{_0}||p_{_1})$ for each $P_x$. Noting that the equality in the constraint $\mathcal{D}(p_{_0}||p_{_1}) \leq 2 \epsilon^2$ for the normal $p(x)$ should be guaranteed, Fig.~\ref{fig:fig3} confirms that the skew-normal $p(x)$ can achieve a higher $I(\mathbf{x},\mathbf{z})$ than the normal $p(x)$ subject to $\mathcal{D}(p_{_0}||p_{_1}) \leq 2 \epsilon^2$. {We note that in Fig.~\ref{fig:fig3} the skew parameter $\theta$ is not optimized in terms of maximizing $I(\mathbf{x},\mathbf{z})$ subject to $\mathcal{D}(p_{_0}||p_{_1}) \leq 2 \epsilon^2$. With an optimized $\theta$, which can be numerically obtained, the skew-normal $p(x)$ can possibly achieve a higher $I(\mathbf{x},\mathbf{z})$ for a given $\mathcal{D}(p_{_0}||p_{_1})$. This observation explicitly shows that Gaussian signaling is not optimal for covert communications with $\mathcal{D}(p_{_0}||p_{_1}) \leq 2 \epsilon^2$ as the constraint.}

\begin{figure}[!t]
    \begin{center}
   {\includegraphics[width=3.5in, height=2.9in]{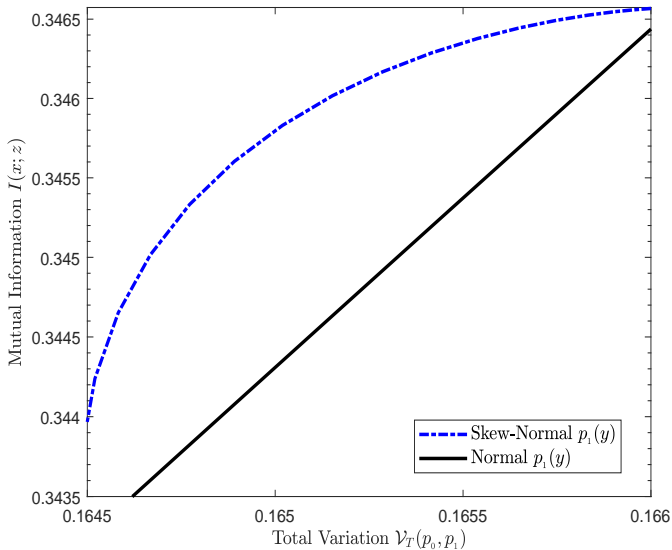}}
    \caption{The mutual information $I(\mathbf{x},\mathbf{z})$ versus the total variation $\mathcal{V}_T(p_{_0}, p_{_1})$ for skew-normal and normal $p_{_1}(y)$.}\label{fig:fig4}
    \end{center}
\end{figure}

Following a similar procedure of obtaining Fig.~3 but replacing the KL divergence $\mathcal{D}(p_{_0}||p_{_1})$ with the total variation $\mathcal{V}_T(p_{_0}, p_{_1})$, we plot the achieved mutual information $I(\mathbf{x},\mathbf{z})$ versus $\mathcal{V}_T(p_{_0}, p_{_1})$ in Fig.~4. From Fig.~4, we observe that the skew-normal $p_{_1}(y)$ can achieve a higher $I(\mathbf{x},\mathbf{z})$ for some specific values of $\mathcal{V}_T(p_{_0}, p_{_1})$ than the normal $p_{_1}(y)$. Noting $\xi^{\ast} = 1 - \mathcal{V}_T(p_{_0}, p_{_1})$, this observation indicates that
Gaussian signalling is not optimal for covert communications with the constraint $\xi^{\ast} \geq 1 - \epsilon$. As discussed in the Introduction, we note that the bounds determined by the KL divergences are still useful, since this total variation $\mathcal{V}_T(p_{_0}, p_{_1})$ can only be numerically determined, while these bounds enable operational covert communication systems in practice through guaranteeing stricter covertness constraints.

\begin{figure}[!t]
    \begin{center}
   {\includegraphics[width=3.5in, height=2.9in]{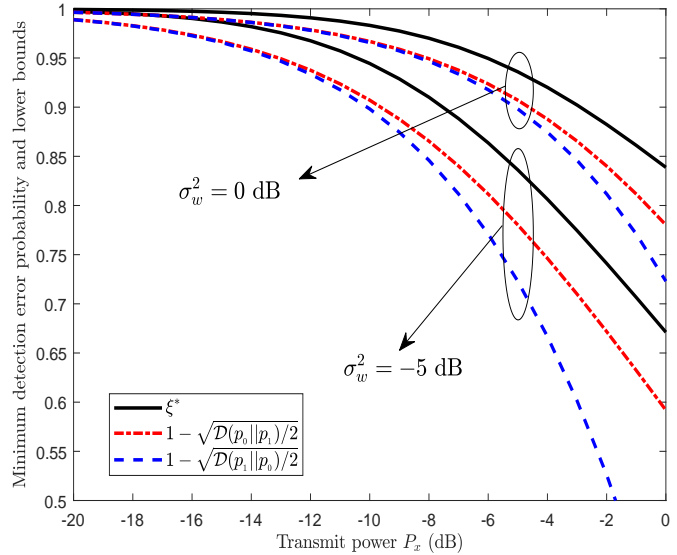}}
    \caption{The minimum detection error probability $\xi^{\ast}$ and its two lower bounds versus the transmit power $P_x$ for different values of $\sigma_w^2$.}\label{fig:fig5}
    \end{center}
\end{figure}

Considering Gaussian signalling, in Fig.~\ref{fig:fig5} we plot the minimum detection error probability $\xi^{\ast}$ and its two lower bounds determined by the two KL divergences, i.e., $\mathcal{D}(p_{_1}||p_{_0})$ and $\mathcal{D}(p_{_0}||p_{_1})$, versus the transmit power $P_x$ for different AWGN power at Willie (i.e., $\sigma_w^2$). In this figure, we first observe that these two lower bounds are close to each other when $\xi^{\ast}$ is close to $1$ for Gaussian signalling. We note that in covertness constraints the value of $\epsilon$ is usually very small, which enforces $\xi^{\ast}$ being close to $1$. This can be the reason why these two bounds have been alternatively used in the literature for covert communications with Gaussian signalling. However, as we have shown in this work, with regard to the optimality of different signalling strategies these two bounds indeed lead two different conclusions.
As expected from our Proposition~\ref{proposition2}, we observe that the lower bound determined by $\mathcal{D}(p_{_0}||p_{_1})$ (i.e., $1 - \sqrt{\mathcal{D}(p_{_0}||p_{_1})/2}$) is tighter than that determined by $\mathcal{D}(p_{_1}||p_{_0})$ (i.e., $1 - \sqrt{\mathcal{D}(p_{_1}||p_{_0})/2}$). Finally, Fig.~\ref{fig:fig5} confirms that $\xi^{\ast}$ decreases as $P_x$ increases or $\sigma_w^2$ decreases.


{
\begin{table}[b]
\caption{{\normalsize Summary of our main results}}
\label{tab:time}%
\centering
\begin{tabular}{c||c|c|c} \hline
    \textbf{Covertness } & \textbf{Gaussian} & \textbf{Constraint}  & \textbf{Maximum} \\
    \textbf{Constraints} & \textbf{Optimality} & \textbf{Strictness}  & \textbf{$I(x; z)$} \\
    \hline \hline
    $\xi^\ast \geq 1 - \epsilon$ & No & Benchmark &  Benchmark   \\
    $\mathcal{D}(p_{_0}||p_{_1}) \leq 2\epsilon^2$ & No & Stricter&  Lower  \\
    $\mathcal{D}(p_{_1}||p_{_0}) \leq 2\epsilon^2$ & Yes & Strictest &  Lowest   \\ \hline
\end{tabular}
\end{table}}

\begin{figure}[!t]
    \begin{center}
   {\includegraphics[width=3.5in, height=2.9in]{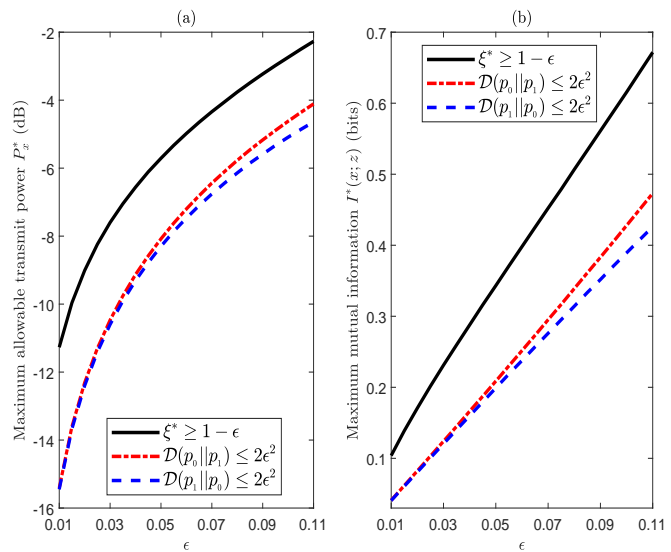}}
    \caption{The maximum allowable transmit power $P_x^{\ast}$ and the maximum mutual information $I(\mathbf{x},\mathbf{z})$ achieved subject to three different covertness constraints, where $\sigma_b^2 = \sigma_w^2 = 0$dB.}\label{fig:fig6}
    \end{center}
\end{figure}

With Gaussian signalling, in Fig.~\ref{fig:fig6} we plot the maximum allowable transmit power $P_x^{\ast}$ and the maximum mutual information $I(\mathbf{x},\mathbf{z})$ achieved subject to three different covertness constraints, i.e., $\xi^{\ast} \geq 1 - \epsilon$, $\mathcal{D}(p_{_0}||p_{_1}) \leq 2 \epsilon^2$, and $\mathcal{D}(p_{_1}||p_{_0}) \leq 2 \epsilon^2$, versus $\epsilon$. In this figure, we first observe that the achieved $P_x^{\ast}$ and the maximum $I(\mathbf{x},\mathbf{z})$ subject to $\xi^{\ast} \geq 1 - \epsilon$ are higher than those achieved subject to the other two constraints. This is due to the fact that $1 - \sqrt{\mathcal{D}(p_{_0}||p_{_1})/2}$ and $1 - \sqrt{\mathcal{D}(p_{_1}||p_{_0})/2}$ are lower bounds on $\xi^{\ast}$, and as shown in Fig.~\ref{fig:fig5} there are gaps between $\xi^{\ast}$ and the two lower bounds. This observation indicates that these two lower bounds are not very tight even in the low regime of $\xi^{\ast}$ for Gaussian signalling, which motivates us to find other tighter lower bounds in future works.
We also observe that $P_x^{\ast}$ or the maximum $I(\mathbf{x},\mathbf{z})$ achieved subject to $\mathcal{D}(p_{_0}||p_{_1}) \leq 2 \epsilon^2$ is greater than that achieved subject to $\mathcal{D}(p_{_1}||p_{_0}) \leq 2 \epsilon^2$. This concludes that $\mathcal{D}(p_{_1}||p_{_0}) \leq 2 \epsilon^2$ is a stricter covertness constraint than $\mathcal{D}(p_{_0}||p_{_1}) \leq 2 \epsilon^2$. As we discussed following our Proposition~\ref{proposition2}, this conclusion holds not only for Gaussian signalling but also for the optimal signalling strategies.

{Following our above examinations, we summarize our main results obtained in this work in Table~I with detailed clarifications. With regard to the results in the second column, Gaussian signalling is not optimal for covert communications with  $\xi^\ast \geq 1 - \epsilon$ as the constraint (as numerically shown in Fig.~4), Gaussian signalling is optimal for covert communications with  $\mathcal{D}(p_{_1}||p_{_0}) \leq 2\epsilon^2$ as the constraint (as proved in Theorem~1), and Gaussian signalling is not optimal for covert communications with  $\mathcal{D}(p_{_0}||p_{_1}) \leq 2\epsilon^2$ as the constraint (as proved in Theorem~3). For the strictness of the covertness constraints as listed in the third column, as we proved in Corollary~1 the covertness constraint $\mathcal{D}(p_{_1}||p_{_0}) \leq 2\epsilon^2$ is relatively stricter than $\mathcal{D}(p_{_0}||p_{_1}) \leq 2\epsilon^2$, since the solution to the optimization problem (15) with $\mathcal{D}(p_{_1}||p_{_0}) \leq 2\epsilon^2$ as the covertness constraint  is feasible to the optimization problem (34) with $\mathcal{D}(p_{_0}||p_{_1}) \leq 2\epsilon^2$ as the covertness constraint. This is due to the fact that for Gaussian signalling we have $\mathcal{D}(p_{_0}||p_{_1}) \leq \mathcal{D}(p_{_1}||p_{_0})$ and Gaussian signalling is the solution to  the optimization problem (15). Since the two KL divergences only determine two lower bounds on $\xi^\ast$, both $\mathcal{D}(p_{_1}||p_{_0}) \leq 2\epsilon^2$ and $\mathcal{D}(p_{_0}||p_{_1}) \leq 2\epsilon^2$ are stricter covertness constraints relative to $\xi^\ast \geq 1 - \epsilon$. The results listed in the fourth column are achieved as per those detailed in the third column accordingly. We note that, although we have proved $\xi^{\ast} \geq 1\!-\!\sqrt{\mathcal{D}(p_{_0}||p_{_1})/2} \geq 1\!-\!\sqrt{\mathcal{D}(p_{_1}||p_{_0})/2}$ for Gaussian signalling in Corollary~2, we cannot draw any conclusion on the tightness of the two bounds (i.e., $1\!-\!\sqrt{\mathcal{D}(p_{_0}||p_{_1})/2}$ and $1\!-\!\sqrt{\mathcal{D}(p_{_1}||p_{_0})/2}$) on the minimum detection error probability $\xi^{\ast}$, since the relationship between $\mathcal{D}(p_{_0}||p_{_1})$ and $\mathcal{D}(p_{_1}||p_{_0})$ has not been clarified for general signalling strategies.}

\section{Conclusion}\label{conclusion}

In this work, we first proved the optimality of Gaussian signalling for covert communications with $\mathcal{D}(p_{_1}||p_{_0}) \leq 2 \epsilon^2$ as the constraints. To this end, we proved that Gaussian signalling can minimize the KL divergence $\mathcal{D}(p_{_1}||p_{_0})$ while maximizing the mutual information $I(\mathbf{x},\mathbf{z})$ subject to power constraints. Unexpectedly, we also proved that Gaussian signalling is not optimal for covert communications with $\mathcal{D}(p_{_0}||p_{_1}) \leq 2 \epsilon^2$ as the constraint, for which the optimal signalling will be tackled in our near future works. As we showed, a skew-normal $p(x)$ can achieve a higher $I(\mathbf{x},\mathbf{z})$ subject to $\mathcal{D}(p_{_0}||p_{_1}) \leq 2 \epsilon^2$ than the normal $p(x)$. Furthermore, as we proved $\mathcal{D}(p_{_1}||p_{_0}) \leq 2 \epsilon^2$ is stricter than $\mathcal{D}(p_{_0}||p_{_1}) \leq 2 \epsilon^2$ as the covertness constraint, which is due to $\mathcal{D}(p_{_0}||p_{_1}) \leq \mathcal{D}(p_{_1}||p_{_0})$ for Gaussian signalling and leads to the fact that $\mathcal{D}(p_{_0}||p_{_1}) \leq 2\epsilon^2$ gives higher mutual information than $\mathcal{D}(p_{_1}||p_{_0}) \leq 2\epsilon^2$.



\end{document}